\documentclass[a4paper,11pt]{article}

\pdfoutput=1

\usepackage{adjustbox}  
\usepackage{booktabs}
\usepackage[T1]{fontenc}
\usepackage[utf8]{inputenc}
\usepackage{jcappub}
\usepackage{url}

\title{Non-linear Structure Formation for Dark Energy Models with a Steep Equation of State}

\author[a,b,c]{N. Chandrachani Devi,}
\author[b,d]{M. Jaber-Bravo,}
\author[a]{G. Aguilar-Arg\"uello,}
\author[a]{O. Valenzuela,}
\author[b]{A. de la Macorra,}
\author[a]{and H. Vel\'azquez}

\affiliation[a]{Instituto de Astronom\'ia, Universidad Nacional Aut\'onoma de M\'exico, A. P. 70-264, 04510, M\'exico, D.F., M\'exico\\}
\affiliation[b]{Instituto de F\'isica, Universidad Nacional Autonoma de M\'exico, Circuito de la Investigación Científica Ciudad Universitaria, 04510, M\'exico, D.F., M\'exico\\}
\affiliation[c]{Institute for Computational Cosmology, Department of Physics, Durham University, South Road, Durham, DH1 3LE, UK\\}
\affiliation[d]{
Institute for Astronomy, Faculty of Physics, Astronomy and Informatics,  \\Nicolaus Copernicus University, Grudziadzka 5, 87-100 Toru\'n, Poland
}

\emailAdd{chandrachani@gmail.com}
\emailAdd{jaber@astro.umk.pl}
\emailAdd{gaguilar@astro.unam.mx}
\emailAdd{octavio@astro.unam.mx}
\emailAdd{macorra@fisica.unam.mx}
\emailAdd{hmv@astro.unam.mx}

\abstract{We study the nonlinear regime of large scale structure formation considering a dynamical dark energy (DE) component determined by a Steep Equation of State parametrization (SEoS) $w(z)=w_0+w_i\frac{(z/z_T)^q}{1+(z/z_T)^q}$. In order to perform the model exploration at low computational cost, we modified the public code L-PICOLA. We incorporate the DE model by means of the first and second order matter perturbations in the Lagrangian frame and the expansion parameter. We analyse deviations of SEoS models with respect $\Lambda$CDM in the non-linear matter power spectrum ($P_k$), the halo mass function (HMF),and the two point correlation function (2PCF). On quantifying the nature of steep (SEoS-I) and smooth transitions in DE field (CPL-lim), no signature of steep transition is observed, rather found the overall impact of DE behaviors in $P_k$ at level of $\sim 2-3\%$ and $\sim 3-4\%$ differences w.r.t $\Lambda$CDM at $z=0$ respectively. HMF shows the possibility to distinguish between the models at the high mass ends.  
The best fitted model assuming only background and linear perturbations dubbed as SEoS-II largely deviates from $\Lambda$CDM and current observations on studying the non linear growth. This large deviation in SEoS-II also quantified the combined effect of the dynamical DE and the larger amount of matter contained, $\Omega_{m0}$ and $H_{0}$ accordingly. 2PCF results are relatively robust with $\sim 1-2 \%$ deviation for SEoS-I and CPL-lim and a significant deviation for SEoS-II throughout $r$ from $\Lambda$CDM. Finally, we conclude that the search of viable DE models (like the SEoS) must include non-linear growth constraints.
}

\begin{document}
\maketitle
\flushbottom

\section{Introduction}
\label{sec:intro}

The unknown physics behind the observed late time accelerated expansion \cite{Roukema:1993yra, Riess:1998cb,Perlmutter:1998np} of our Universe has motivated the scientific community to join their efforts in developing various spectroscopic and imaging galaxy surveys. 
These include the extended Baryon Oscillation Spectroscopic Survey (eBOSS)\footnote{https://www.sdss.org/surveys/eboss/}, the dark energy survey (DES)\footnote{https://www.darkenergysurvey.org/es/}, the Dark Energy Spectroscopic Instrument (DESI)\footnote{https://www.desi.lbl.gov/}, Large Synoptic Survey Telescope-LSST \footnote{https://www.lsst.org/}, the European Space Agency's Euclid \footnote{https://www.euclid-ec.org/}, etc. They aim to measure the cosmic expansion history and the growth of structure to a precision at one percent-level. 
These huge upcoming datasets will open a window to perform several tests to discriminate between various competing cosmological models in the near future (see \cite{Santos:2016sti,Santos:2016sog,Lonappan:2017lzt} for some studies on discriminating models). 

So far, the concordance $\Lambda$ cold dark matter 
cosmological model ($\Lambda$CDM) remains as  the most successfully tested and consistent model with several observations. However, this model suffers from deep theoretical issues namely the cosmic coincidence and the fine tunning problems, which, for instance, refers to having a very small magnitude in comparison to the expected value from the fundamental physics \cite{Weinberg1989,Sahni:1999gb}. Beyond the $\Lambda CDM$ model, a number of alternative theories have been proposed in the literature to explain this late time cosmic acceleration \cite{Copeland:2006wr,DeFelice:2010aj,Clifton:2011jh,Joyce:2014kja}. In general, these alternative theories, either consider an exotic matter component with a large negative pressure called dark energy (DE) \cite{Copeland:2006wr} or modify the Einstein theory of Gravity (GR) on cosmological scales \cite{Koyama2016, Clifton:2011jh, Joyce:2014kja, Jaime:2012gc}. 

Recently, various observational anomalies have arisen in the context of the $\Lambda$CDM cosmology \cite{Raveri:2015maa,Macaulay:2013swa}. For instance, the tension between the local value measurement of the Hubble parameter, $H_0$ and the one extrapolated from CMB measurements assuming the cosmological constant \cite{Aghanim:2018eyx}. It has reached upto the $5.3\sigma$ level \cite{Wong:2019kwg}. There is also a well known tension in the growth factor measurement, $f\sigma_{8}$, between the value measured from galaxies clustering (e.g. redshift-space distortions) and the value predicted by Planck Collaboration \cite{Macaulay:2013swa}. A plausible explanation may come from the correct determination of different systematic errors on those experiments. However, such tensions could also be a hint to new physics beyond the concordance $\Lambda$CDM.

On the other hand, studies of galaxies and dark matter clustering turn out to be a promising probe, not only with the possibility to break down the degeneracy between different DE models but also to test different Modified Gravity scenarios and their screening mechanisms (see for references \cite{Devi2019, Hernandez-Aguayo:2018oxg, Hernandez-Aguayo:2018yrp}). With the upcoming high precision galaxies surveys, an exploration of alternative models beyond the background expansion into both quasi-linear and non-linear regimes has become one of the important and necessary tasks in modern cosmology. In this regards, N-body simulations play a vital role along with the perturbative approaches in order to trace accurately the growth of structures particularly for the non-linear regimes.

There are some N-body simulations available based on dynamical DE, such as \cite{Lawrence+2017,Garrison:2017ssz, Almaraz:2019zxy}. The halo catalogs of 125 cosmological N-body simulations under the Abacus project are released in \cite{Garrison:2017ssz} where 40 sets are based on the constant equation of state DE ($wCDM$) model.
 While \cite{Lawrence+2017} provided the results of the Chevallier-Polarski-Linder parameterization \cite{Chevallier2001,Linder2003} CPL-parameterised DE model including the massive neutrinos. The later were run with the Hardware-Hybrid Cosmology Code (HACC), one of a high-performance cosmology code (see for more details in \cite{Heitmann:2015xma}).

In addition to the high resolution N-body simulations, various fast approximated N-body simulation have been developed with a motive to accurately mock the galaxy catalogs on vast scales. 
However, their primary goal is to estimate accurately the covariance matrix for concerned surveys but they are based on the standard $\Lambda$CDM model. 
The lists include PTHalos \cite{Scoccimarro:2001cj}, Quick Particle Mesh Simulations(QPM) \cite{White:2013psd}, Effective Zel'dovich approximation mocks(EZmocks) \cite{Chuang:2014vfa}, PINOCCHIO \cite{Monaco:2001jg, Monaco:2013qta}, PATCHY \cite{Kitaura:2014mja}, COmoving Lagrangian Acceleration-COLA \cite{Tassev:2013pn,Tassev:2015mia}, an upgraded light cone enable parallel version of COLA: L-PICOLA \cite{Howlett:2015hfa}, etc.  Each of them has its own approximation methods and applications on mocking the catalogs and analyzing uncertainties on several on-going surveys. 

 For our purpose, we will employ the L-PICOLA code\cite{Howlett:2015hfa} after incorporating the dynamical DE through the expansion history, first and second order Lagrangian Perturbation Theory (2LPT) approximation. The COLA method used in \cite{Howlett:2015hfa} was introduced with an idea to capture the large scale structure(LSS) accurately within few time steps, with the less used of computational resources but still enabled to trace accurately the small scales \cite{Tassev:2013pn,Tassev:2015mia}. It is done successfully by implementing the solutions of the first and second order dark matter perturbations that guarantee to capture the LSS to some extend. Along with this idea of COLA and the large number of cosmological models we have, an implementation of such models into this fast COLA-like N-body simulations would definitely be an efficient and adequate way to understand their effects in the regime required for the upcoming galaxy surveys.
 
An attempt in this regard has already made for the modified gravity(MG) theories in the referred code called MG-COLA \cite{Winther:2017jof}.
Their studies pointed out that in general the COLA approach overestimates the halo mass function even for the $\Lambda$CDM model but preserving the difference of MG with respect to $\Lambda$CDM accurately. For our case, we are choosing the dynamical DE models of CPL like parameterization \cite{Chevallier2001,Linder2003} and the steep equation of state parameterization \cite{Jaber:2017bpx}. A comparative study will be performed with respect to $\Lambda$CDM. The main idea of the work is not only to understand the effect of these DE models but also to confront our understanding in disentangling their effect from the global cosmological parameters such as $\Omega_{m0}$ and $H_{0}$, through LSS.

The paper is structured as follows: in Section \ref{sec:method}, we describe the models, their background evolution and the dark matter perturbations up to second order (see for a review \cite{Bernardeau:2001qr}). In Section \ref{sec:nonlinear}, we discuss the COLA method that employed by the N-body simulation called L-PICOLA\cite{Howlett:2015hfa}. The results in term of the dark matter power spectrum, the halo mass function and two point correlations function of halos are discussed in section \ref{sec:results}. We summarise and conclude in Section \ref{sec:conclusions}.

The fiduciary cosmological parameters used for this work are $\Omega_{m0} = 0.3089$, $\Omega_{b0} = 0.045$, $h = 0.6774$, $\sigma_{8} = 0.8159$,and $n_s = 0.9667$.

\section{Cosmological models}
\label{sec:method}
This section focuses on the DE models we considered for our analysis. Their background expansion history, and the evolution of their matter perturbations up to second order are discussed. Such perturbative solutions are used in generating the initial conditions for the N-body simulations.

\subsection{Dark energy model and background expansion}
\label{sec:background}

\begin{figure}
\centering 
\includegraphics[width=0.6\textwidth, angle=0]{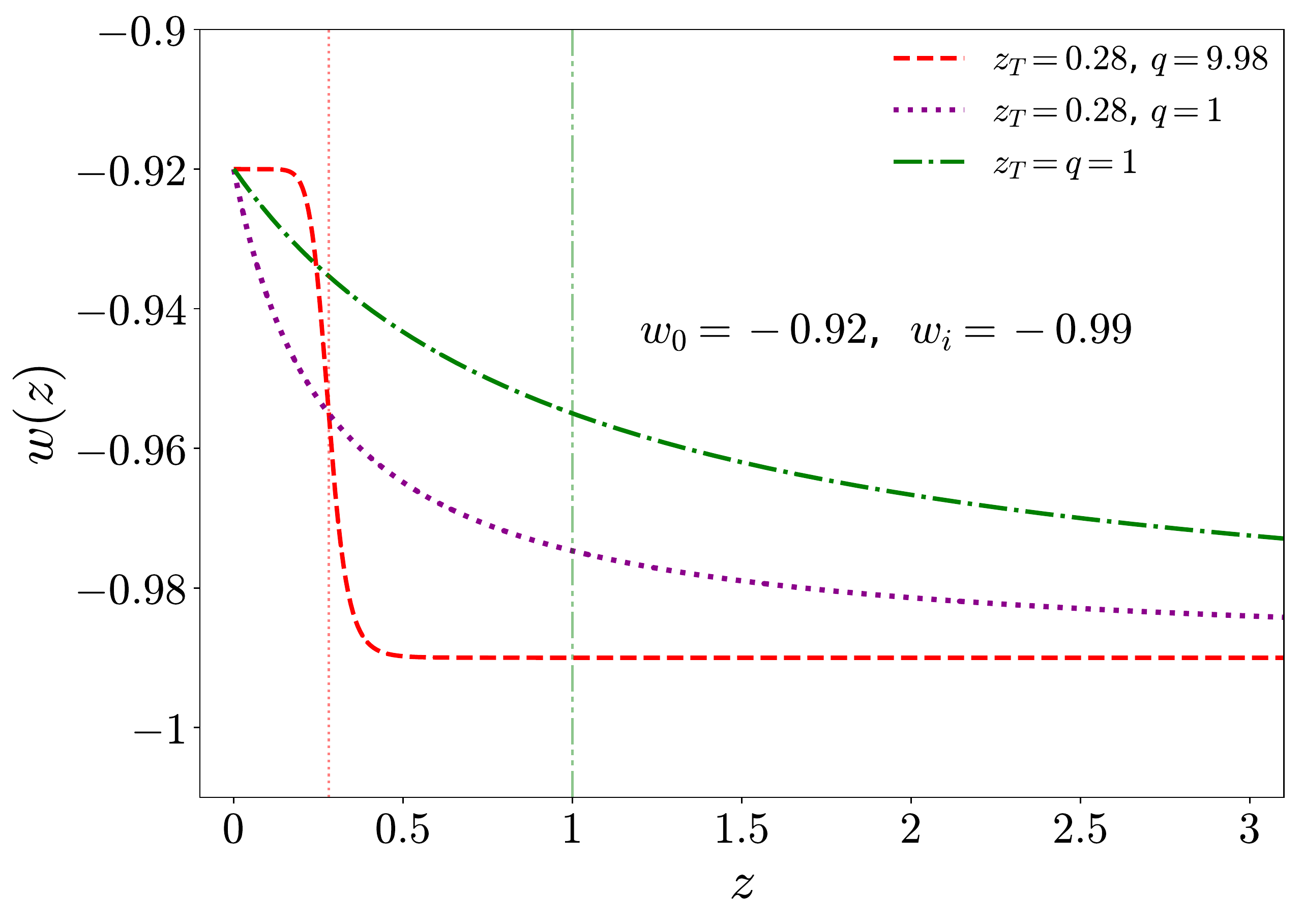}
\caption{\label{fig:eosqs} Evolution of equation of state parameter (EoS) for the steep DE models with $w_0=-0.92$, $w_i=-0.99$. The red dashed line corresponds to $z_T=0.28$ and $q=9.98$ which we refer as SEoS-I afterwards. The green dot-dashed line with $q=z_T=1$ represents the CPL limit of SEoS model with the CPL parameters: $w_0 =-0.92$ and $w_a =-0.07$. The magenta dotted line with $z_T=0.28$ and $q=1$ represents an intermediate case between them. The transition between two pivotal points is marked with the vertical lines for SEoS-I and CPL limit that are occurring at $z_T=0.28$  and $z_T=1$, respectively. The value $z_T$ is defined  as $z$ where $w(z_T) = (w_0+w_i)/2$.
}
\end{figure}

In a phenomenological approach, dark energy models are parametrized through its equation of state (EoS), $w(z) \equiv \frac{{\rm p_{DE}}(z)}{{\rm \rho_{DE}}(z)}$, which describes the evolution of DE pressure to its density as a function of time within a set of free parameters. 
Several forms of $w(z)$ have been proposed in the literature, (see for instance \cite{Chevallier:2000qy, Linder2003, Doran:2006kp, KraussJonesHuterer2007, Linder:2006ud, Rubin:2008wq, 2009ApJ:703:1374S, 2010PhRvD..81f3007M, Hannestad:2004cb, Jassal:2004ej, Ma:2011nc, Huterer:2000mj, Weller:2001gf, Huang:2010zra, Barboza:2008rh}). 
 Among them, the so-called Chevallier-Polarski-Linder (CPL) parameterization \cite{Chevallier2001,Linder2003} is one of the most widely studied parameterization to the DE equation of state and is given by:
\begin{equation}
w(z) =w_{0}+w_{a}\frac{z}{1+z},
\label{eq:wcpl}
\end{equation}
where $w_{0}$ and $w_{a}$ are constant parameters representing the present value of EoS and its overall time evolution, respectively. Current observational constrains on CPL parameters are presented in \cite{Linden:2008mf}. 

Within the framework of General Relativity, for a flat, homogeneous and isotropic background universe (i.e. the universe governed by the FLRW metric) which consists of the non-relativistic dark matter (DM) and the DE component i.e. $\Omega_{DE} + \Omega_{m} = 1$, the Hubble expansion of our universe is governed by
\begin{equation}
\frac{H(z)^2}{H_{0}^2} = \Omega_{m0} (1+z)^{3} + (1- \Omega_{m0}) F(z),
\label{eq:hubble}
\end{equation}
where $H_{0}$ and $\Omega_{m0}$ represent the present day values of Hubble rate and matter density respectively. The DE density, $\Omega_{DE}(z) = (1-\Omega_{m0})F(z)$ evolves with redshift, $z$ as 
\begin{equation}
F(z) = \exp\left[3 \int_{0}^{z} (1+w(z^{\prime}))/(1+z^{\prime})dz^{\prime}\right].
\label{eq:CPL}
\end{equation}
In this work, we consider a more general EoS of DE other than \eqref{eq:wcpl}, inspired by  quintessence fields dynamics \cite{delaMacorra:2015aqf}, a parametrization described by:
\begin{equation}
\label{eq:seos}
w(z) = w_0 + (w_i-w_0)\frac{\left(z/z_T \right)^q}{1+\left(z/z_T \right)^q}.
\end{equation}
where $w_i$ and $w_0$ represent the value of $w(z)$ at high redshifts and at present day, respectively. Here the function, $f(z) = \frac{\left(z/z_T \right)^q}{1+\left(z/z_T \right)^q}$ with a transition redshift, $z_{T}$, and a steepness parameter, $q$, governs the dynamics of the parametrization between two pivotal values ($w_i$,$w_0$). For instance, in case of $f(z) =0$ at $z=0$, EoS becomes $w(z) = w_{0}$; at $ z -> \infty$ where $f(z) =1$, $w(z) = w_{i}$ and at $z =z_T$ where $f(z) = 1/2$, then $w(z) = (w_0+ w_i)/2$. This gives us the range of $f(z)$ values, i.e $0 < f(z) < 1$. 

Note that $q$ determines the steepness of transition. With a large value of $q$, an abrupt transition with a shorter period of transition is expected and vice-versa. This feature derives the name of parametrization as “Steep Equation of State” DE models, afterward as SEoS models. Interestingly, the well-known CPL model (\ref{eq:CPL}) is recovered with a smooth transition at $z_T = 1$ for $q=1$, where the CPL parameters relate with SEoS parameters via $w_a\equiv w_i-w_0$.
We further consider this particular case with ($w_0 = -0.92$, $w_i = -0.99$) and $z_T=q= 1$ as our “CPL limit” for comparison of different type of DE models.
The model also recovers the standard $\Lambda$CDM model in the limit of $w_0 = w_i = -1$ for any values of $q$ and $z_T$.


Figure (\ref{fig:eosqs}) shows the evolution of $w(z)$ of SEoS models with its free parameters fixed to the best fit values from the previous studies \cite{Jaber:2017bpx}: $\{w_0, w_i, q, z_T\} = \{-0.92,-0.99,9.98,0.28\}$(red dashed line), refer as SEoS-I along with slightly varying values of $(q, z_{T})$. The corresponding CPL limit (CPL-lim) with its parameters: $w_0 =-0.92$ and $w_a =-0.07$ is presented with the green dot-dashed line. For the same values of ($w_0, w_i, z_T$) but with $q=1$, we have a different behavior, shown in the magenta dotted line. This serves as an intermediate case between the above two. The transition redshift, $z_T$ between two pivotal points $(w_0, w_i)$ for SEoS-I and CPL limit are marked by the vertical lines that occur at $z_T=0.28$  and $z_T=1$, respectively. For more details of the EoS behaviors, we refer to figure(1) of \cite{Jaber:2019opg}.

One can see from the figure (\ref{fig:eosqs}) that $w(z)$ with $q=9.98$ (red dashed line) transits from $\sim 0.1\%$ difference w.r.t $\Lambda$CDM to $\sim 8\%$ in the interval of $\Delta z \sim 0.2$, presenting a rapid dilution of the DE density. 
While in CPL-lim (green dash line) with $q = 1$, the transition occurs from $\sim 0.3\%$ to $\sim 8\%$ difference in the interval of $\Delta z \sim 2$, clearly depicting how smooth transition can be depending on the values of $q$. For more details for the EoS behaviors, we refer to figure(1) of \cite{Jaber:2019opg}.

The models we considered are summarised in table (\ref{table:COLA_models}). The SEoS-I and CPL-lim refer to SEoS models with  $\{w_0, w_i, q, z_T\} = \{-0.92, -0.99, 9.98, 0.28\}$ and its CPL limit mentioned above, with the global cosmological parameters based upon the Planck collaboration (P15)\cite{Ade2016} as for $\Lambda$CDM. The SEoS-II model is defined with the same SEoS parameters $\{w_0, w_i, q, z_T\} = \{-0.92,-0.99, 9.98,0.28\}$ but with the best fit values of $H_{0}$ and $\Omega_{m0}$ from \cite{Jaber:2017bpx} i.e. $H_{0}=73.22$ and $\Omega_{m0} =0.334$. According to \cite{Jaber:2017bpx}, the
SEoS-II setting was firmed to be the best fitted model among SEoS type. However, in \cite{Jaber:2017bpx} observations of the Baryonic acoustic peak from galaxies and Lymann-$\alpha$ forest measurements were used along with the local determination of the Hubble parameter $H_0$ from \cite{Riess:2016jrr} to constrain the free parameters of this model. Their analysis found $1\sigma$ constraints on the parameters: $w_0=-0.92^{+0.15}_{-0.14}$, $w_i=-0.99(\leq-0.67)$. The parameters $q$ and $z_T$ were fit to the values $q=9.98$ and $z_T=0.28$ but no constraints were found at 1$\sigma$ level using those data sets.

\begin{figure}
\centering 
\includegraphics[width=0.6\textwidth,angle=0]{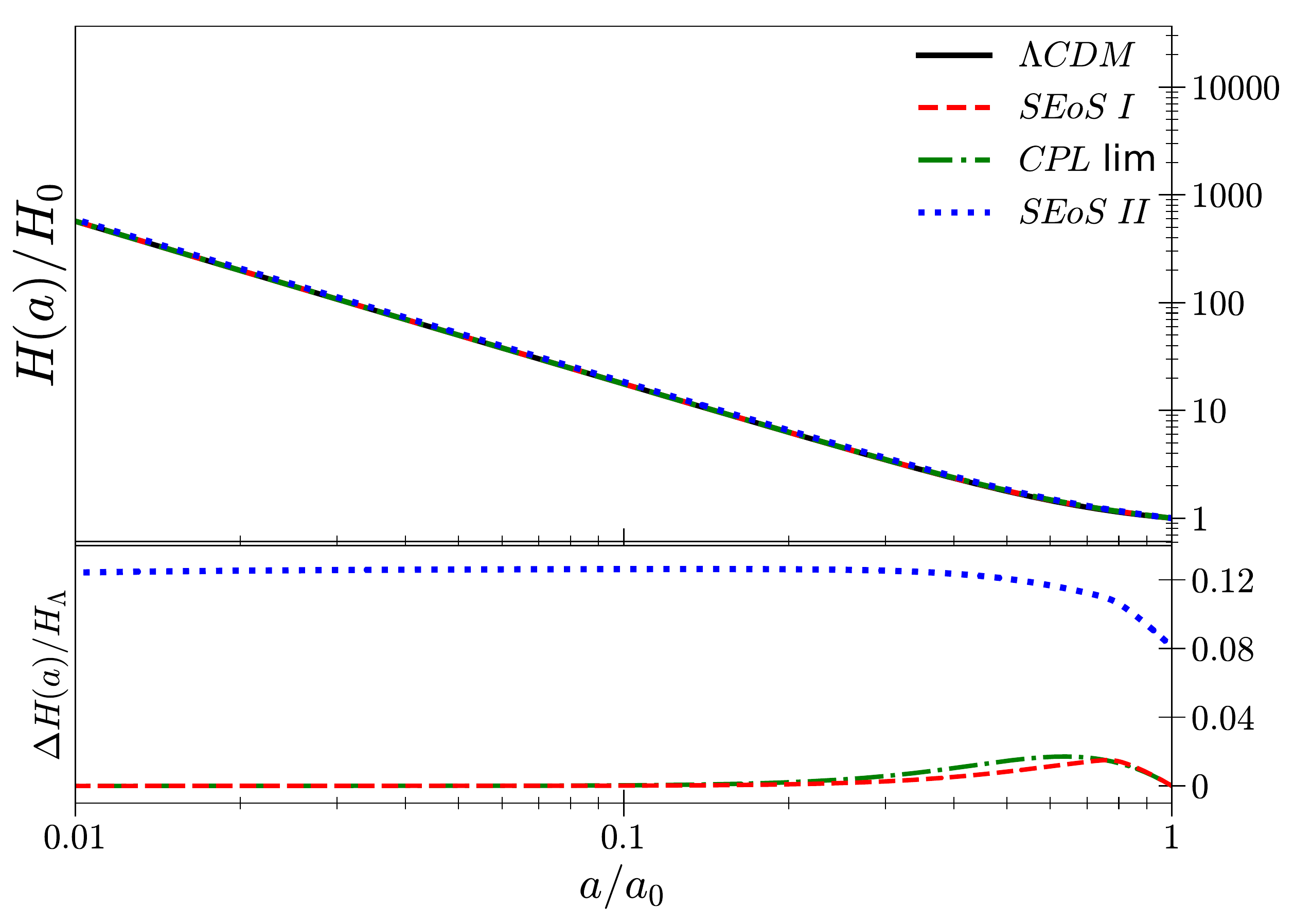}
\caption{\label{fig:hubble_ratios}{\it Upper panel}: Normalized Hubble expansion rate for the models we considered (as listed in Table \ref{table:COLA_models}). {\it Bottom panel}: The relative difference with respect to $\Lambda$CDM: $\Delta H/H_{\Lambda} \equiv (H/H_{\Lambda} -1)$. Since the models SEoS-I (red dash line) and $CPL$-lim (green dotted-dash line) share the same values of $H_0$ and $\Omega_{m0}$, they converge to the same $H(z=0)$ as that of $\Lambda$CDM (black solid line). The CPL limit of SEoS model, with the transition redshift at $z_T=1$ which is larger compare to the SEoS case, $z_T=0.28$ and therefore it shows a departure from $\Lambda CDM$ at a smaller $a/a_0$ value. Since the SEoS-II model evolves with different $H_0$ and also has different amount of matter contained, a discrepancy from $\Lambda CDM$ is observed throughout all time scales. Note: This convention will be followed while presenting these models in the rest of figures.} 
\label{figure:hubble}
\end{figure}

The corresponding Hubble expansion rates are shown in figure (\ref{fig:hubble_ratios}) for the models condensed in table (\ref{table:COLA_models}). The solid black, red dashed, green dotted-dash, blue dotted lines represent the $\Lambda$CDM, SEoS-I, CPL-lim and SEoS-II respectively. This convention will be used representing these models in rest of the paper. 

The lower panel of figure (\ref{fig:hubble_ratios}) shows that SEoS-I and CPL-lim being shared the same cosmological parameters as $\Lambda$CDM, they, thus, converge into the same $H(z)$ value at $z=0$. But in CPL-lim, the transition occurs at higher redshift in comparison to SEoS-I, so it starts to deviate from $\Lambda$CDM slightly earlier then SEoS-I. However, the expansion rate, $E(a) =H(a)/H_{0}$, of SEoS-I and CPL-lim remain within $\sim 1\%$ difference from $\Lambda$CDM at most of the time except they go up to the difference of $\sim 2\%$ around the transitions epoch, $z_{T} = 0.28, 1$. In case of SEoS-II, despite of the same SEoS parameters, due to the different values of $H_0$ and $\Omega_{m0}$, some discrepancies from $\Lambda CDM$ are expected and observed throughout all the redshift values. In particular, we notice that at present day, the value for $H(z)$ is $\sim 8\%$ larger than in $\Lambda CDM$ and the difference goes up to $12\%$ at higher redshifts. 

\begin{figure*}
     \centering
     \begin{tabular}{cc}
        \includegraphics[width=0.49\textwidth,angle=0]{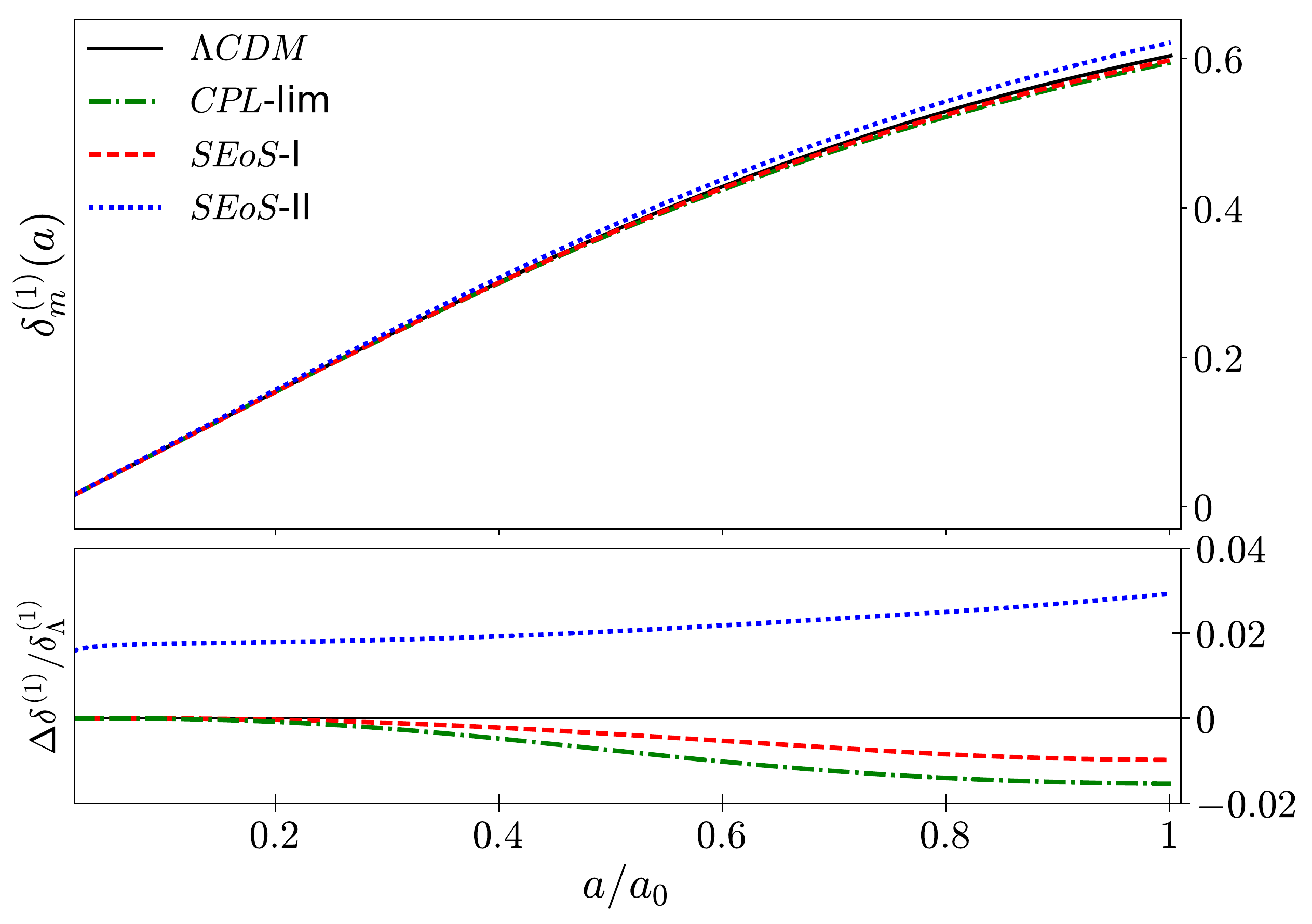}   
        \includegraphics[width=0.49\textwidth,angle=0]{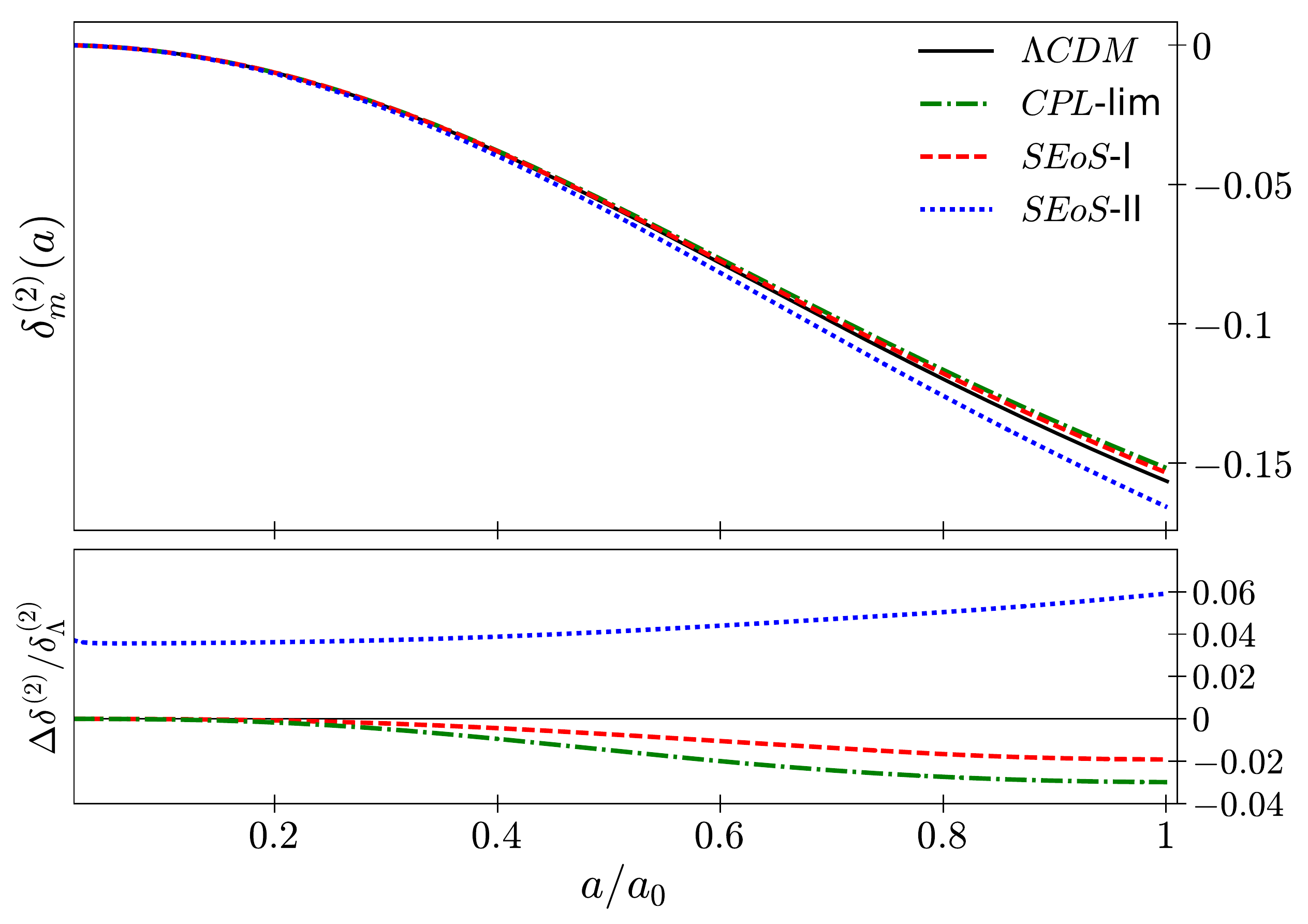} \\

     \end{tabular}

\caption{
Evolution of the dark matter density contrast at first $\delta_m^{(1)}(a)$ and second order, $\delta_m^{(2)}(a)$, for the models considered in the left and right panels respectively. The relative differences w.r.t to $\Lambda$CDM are shown in their respective bottom panels. The SEoS-I turns out be the closest in behavior with $\Lambda$CDM, followed by CPL-lim and SEoS-II. 
}
\label{fig:deltas}
\end{figure*}

\label{sec:linear}

\begin{figure*}
     \centering
     \begin{tabular}{cc}
        \includegraphics[width=0.5\textwidth,angle=0]{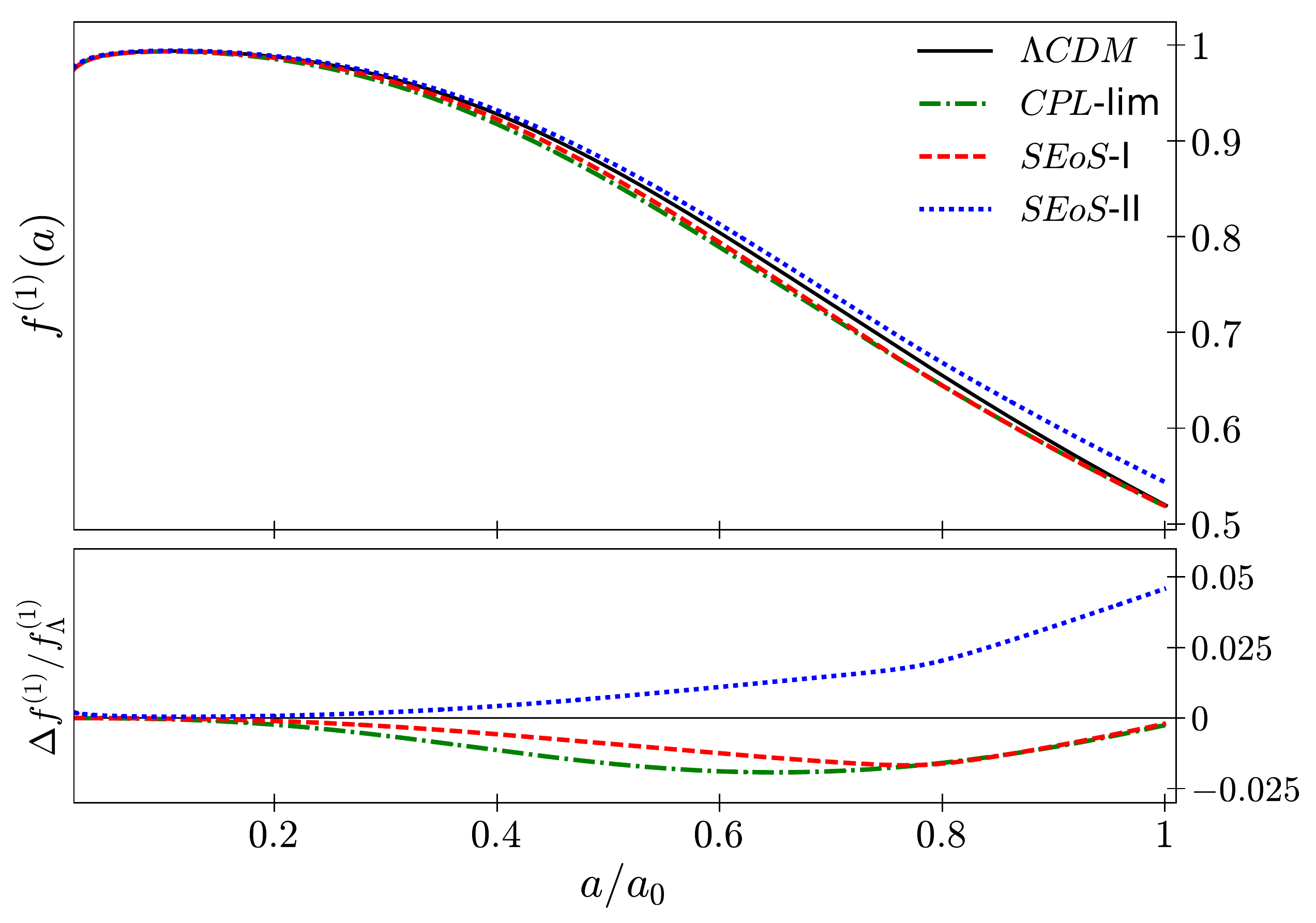}   
        \includegraphics[width=0.5\textwidth,angle=0]{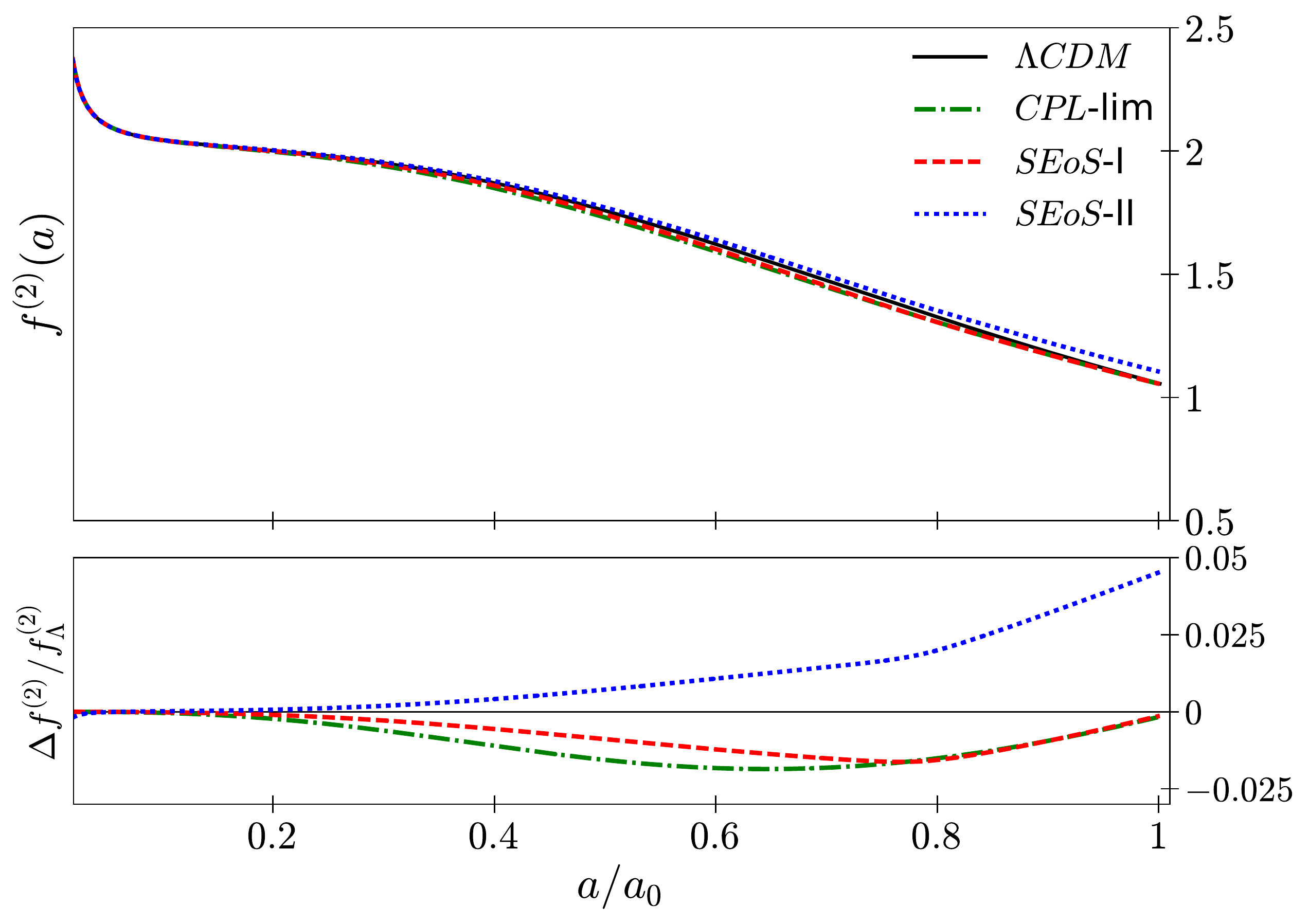} \\

     \end{tabular}

\caption{Evolution of the growth function at first, $f^{(1)}(a)$, and second order, $f^{(2)}(a)$,  for the models we considered in upper-left and upper-right panels, respectively. The relative differences w.r.t to $\Lambda$CDM are showing in the bottom panels. The $SEoS$-II model has comparatively higher growth factor in both orders at low redshift, implying faster growing mode then other models. 
}
\label{fig:f_growth}
\end{figure*}

\subsection{Matter Perturbation Theory}

The evolution of large scale structure are understood to some extent by studying the linear dark matter perturbation theory \cite{Ma:2011nc, Bernardeau:2001qr}. 
Under the assumption that DE remains homogeneously distributed in all scales of our universe, i.e no clustering in the DE field, $\delta \rho_{DE}= 0$ 
, the evolution of the matter density contrast, $\delta_{m} = \frac{\rho_m}{\bar{\rho}_{m}}-1$ up to first and second order are given by:


\begin{eqnarray}
a^2 \frac{d^2\delta_m^{(1)}}{da^2}+\frac{3}{2}\left(2- w(a)\Omega_{DE}(a)\right)\frac{d\delta_m^{(1)}}{da} -\frac{3}{2}\Omega_m(a)\delta_m^{(1)}(a) = 0,\label{eq:2lptDE1} \\
a^2 \frac{d^2\delta_m^{(2)}}{da^2}+\frac{3}{2}\left(2-w(a)\Omega_{DE}(a)\right)\frac{d\delta_m^{(2)}}{da} -\frac{3}{2}\Omega_m(a)\delta_m^{(2)}(a)= -\frac{3}{2}\Omega_m(a)\left[\delta_m^{(1)}(a)\right]^2 ,
\label{eq:2lptDE2}
\end{eqnarray}
where $\Omega_m(a) = \frac{H_{0}^2\Omega_{m0}a^{-3}}{H(a)^2}$ and $\Omega_{DE}(a) =  \frac{H_{0}^2(1-\Omega_{m0})F(a)}{H(a)^2}$. 
The above coupled differential equations are solved by applying the initial conditions at the time around the recombination epoch, $a_{ini} \approx 0.001$, where the universe is in the phase of matter dominated ``Einstein-de-Sitter'' phase, $\Omega_{m} = 1$. In that phase, $\delta^{(1)}_{m}(a)$ evolves linearly with the scale factor and hence we set $\delta_{m}^{(1)}(a_{ini})=a_{ini}$ and  $d\delta_{m}^{(1)}/da(a) = 1$ at $a= a_{ini}$. Likewise, $\delta^{(2)}_{m}(a) \sim -3/7 a^2$ and  $d\delta_{m}^{(2)}/da(a) \sim a $ at $a= a_{ini}$.

Figure (\ref{fig:deltas}) shows the evolution history of $\delta_{m}^{(1)}$ and $\delta_{m}^{(2)}$ obtained from the above equations in the left and right panels respectively. The relative difference w.r.t $\Lambda$CDM are shown in their respective lower panels.
Because of having a slightly higher expansion history in SEoS-I and CPL-lim models compare to $\Lambda$CDM, the matter density contrasts under such scenarios are expected to have lower values w.r.t $\Lambda$CDM. 
Indeed, we observe a comparatively lower matter density contrasts evolution for both SEoS-I and CPL-lim in figure (\ref{fig:deltas}). Particularly, the differences of $\sim 1\%$ to $\sim 2\%$ are shown in $\delta_{m}^{(1)}$ and $\delta_{m}^{(2)}$ for SEoS-I w.r.t $\Lambda$CDM at $z=0$ respectively. The deviation starts to grow around $z \sim 4$ in both cases. In CPL-limit, a deviation reaches to $\sim 1.5\%$ in $\delta_{m}^{(1)}$ and $\sim 3\%$ in $\delta_{m}^{(2)}$ at the same redshift. However, we don't observe any signatures that can manifest the effect of the steep transition in DE density field in these linear perturbation theory studies. 

In case of SEoS-II, even though the Hubble expansion is high compare to $\Lambda$CDM, beings a universe with the high DM contained over other models, the DE epoch comes later and subsequently the matters grow at faster rate that ends into high matter density contrast. Therefore, comparatively $\delta_{m}^{(1,2)}$  are observed to be have larger differences and faster grow over other cases. As expected, the deviation of $\sim 5\%$ are found in $\delta_{m}^{(1,2)}$ at $a = 1$ in SEoS-II. We, thus, speculate that the possibility to distinguish between these dynamical DE models from $\Lambda$CDM is around $3-6\%$ via the second order perturbation theory. 

To quantify the speed of growth of structure under the considered cosmological scenarios, we report the logarithmic growth function at first and second order, $f^{(1,2)}\equiv\frac{d\ln\delta_m^{(1,2)}(a)}{d\ln a}$, in figure \ref{fig:f_growth}. These quantities relate the matter density to velocity dispersion. Figure (\ref{fig:f_growth}) shows that SEoS-I and CPL-lim have comparatively slower growth rates then $\Lambda$CDM in both orders. The rates of growth of structures in SEoS-I and CPL-lim models compare to $\Lambda$CDM have some specific trends of showing down in rate upto a certain scale then start to increase until they reach to $\Lambda$CDM case. The SEoS-I turns out to be the closest one to $\Lambda$CDM and SEoS-II has the fastest growing modes over other models. We do expect to observe some effect in the non-linear regime of structure formations as well.

We also fitted $f^{(1,2)}$ for all models using the known fitting formulae proposed in \cite{Wang:1998gt} that defined $f^{(1,2)}$ as $f^{(1)}=\Omega_m(a)^{\gamma_1}$ and $f^{(2)}= b\times\Omega_m(a)^{\gamma_2}$ with the growth indexes: $\gamma_{1}$, $\gamma_{2}$ and the amplitude parameter, $b$.
The best fitted values of $\gamma_{1}$, $\gamma_{2}$ and $b$ for all models are listed in table \ref{table:COLA_models}. 
Following \cite{Bouchet1994}, we set $\gamma_1=6/11$, $\gamma_2=0.55$ and $ b=1$ for the $\Lambda$CDM model.

%

%



\section{N-body Simulation: Non-linear Evolution}
\label{sec:nonlinear}

\begin{table*}
	\begin{center}
		\begin{tabular}{l l c c c c c c c c c c}
			\hline 
			 Model & Alias &$w_0$ & $w_i$ & $q$ & $z_T$ & $\Omega_{m0}$ & $H_0$ & $n_{s}$ &
			 $\gamma_1$ & $b$ & $\gamma_2$ \\
			 \hline \hline 
			$\Lambda CDM$ & $\Lambda$CDM & -1 & 0 & 0 & 0 & 0.3089 & 67.74 & 0.9667 & 5/9 & 2 & 6/11\\
			\hline
			SEoS  & SEoS-I & -0.92 & -0.99 & 9.98 & 0.28 & 0.3089 & 67.74 & 0.9667 & 0.5527 & 2.0743 & 0.5912\\
			CPL  & CPL lim & -0.92 & -0.99 & 1 & 1 & 0.3089 & 67.74 & 0.9667 & 0.5535 & 2.0751  & 0.5927\\
		\hline
			SEoS & SEoS-II & -0.92 & -0.99 & 9.98 & 0.28 & 0.3340 & 73.22 & 0.9667 & 0.5533 & 2.0859 & 0.5936\\
			\hline 
		\end{tabular}
		\caption{Cosmological and model parameters specification. $\gamma_1$, $\gamma_2$, and $b$, refer to the best fit values found using the fitting formulae for growth factors: $f^{(1)}=\Omega_m^{\gamma_1}$, and $f^{(2)}=b \Omega_m^{\gamma_2}$ \cite{Wang:1998gt}.} 
\label{table:COLA_models}
	\end{center}
\end{table*}

\begin{table*}
\centering
\begin{tabular}{c|c|c}
\hline
parameter & definition & value \\
\hline
\hline
$L_{\rm box}$ & simulation box size & 1024~$h^{-1}$Mpc\\
$N_{\rm p}$ & simulation particle number & $1024^3$\\
$N_{\rm Mesh}$ & FFT Mesh & $1024^3$\\
$z_{ini}$ & Initial redshift & $49$\\
$n_{step}$ & Time steps & $200$\\
$ z_{\rm out}$ & Snapshots out at $z$ & $ 0, 0.1, 0.28, 0.56, 1, 1.5, 2, 2.3, 19 $\\
$ N_{\rm run}$ & Number of run & 5 \\
\hline
\end{tabular}
\caption{Specifications of the N-body simulations.}
\label{table:simulation_COLA}
\end{table*}

In order to understand the structure formation of our Universe in the non-linear regimes, one way is to follow the formation and distribution of DM halos. This can be done through the full N-body DM simulations. However, in order to achieve a reasonable approximation of the structure formation both at large and small scales, the N-body codes need to run numerous time-steps with a fair amount of computational resources.


For our studies, we modify the publicly available N-body DM simulation so-called the COmoving Lagrangian Acceleration method\cite{Tassev:2013pn}, paricularly L-PICOLA\cite{Howlett:2015hfa} for simulating the cosmological structure formation under the dynamical DE models described in section \ref{sec:background}. 
This scheme has an advantage to recover the large scale structures (LSS) accurately in few time steps with less usage of computational resources and is also able to trace accurately the small scales, 
 thanks to the implementation of Lagrangian perturbation theory (LPT), which allows to capture the large scale dynamics directly via the linear growth factors. Note that this makes the code 3 times faster in comparison to the standard N-body code by enabling it to take larger time steps.

 
In the following subsection, we briefly describe the COLA method and the Lagrangian perturbation theory employed in the code.

%

\subsection{COLA Method}


In general, the standard cold-dark matter N-body simulations are governed by a system conformed by the equation of motion and the Poisson equation:
\begin{equation}
\frac{d^2 {\bf x}}{d\tau^2} = - {\bf \nabla} \Phi_{N},
\label{eq:geodesic}
\end{equation}
\begin{equation}
\nabla^2\Phi_{N} = 4 \pi G {\bar{\rho}_{m}}a^4 \delta_{m},   
\label{eq:poisson}
\end{equation}
where $d\tau = \frac{dt}{a}$, $\delta_{m}=\frac{\rho_{m}}{\bar{\rho}_{m}}-1$ and $\Phi_N$ represent the conformal time scale, the matter density contrast corresponding to the particle positions, $\bf{x}$ and the Newtonian potential respectively.

\subsubsection{2LPT}
According to the Lagrangian perturbation theory (LPT) \cite{Bernardeau:2001qr}, the position of a particle in Eulerian space $\bf{x}$ is described by its initial Lagrangian position $\bf{q}$ and a displacement field of the particle, $\bf{\Psi}$ as 
\begin{equation}
\bf{x}(q,\tau) = \bf{q}+\bf{\Psi}(\bf{q},\tau),
\label{eq:LPT_position}
\end{equation}
and the equation of motion (\ref{eq:geodesic}) as given by
\begin{equation}
\frac{d^2 \bf{\Psi}}{d\tau^2} + \mathcal{H}(\tau)\frac{d\bf{\Psi}}{d\tau} +{\bf \nabla}\Phi_N = 0.
\label{eq:geodesic2}
\end{equation}
Hence, the Poisson equation (\ref{eq:poisson}) can be written as
\begin{equation}
{\bf \nabla_{x}}. \left( \frac{d^2 \bf{\Psi}}{d\tau} + \mathcal{H}(\tau)\frac{d\bf{\Psi}}{d\tau}\right) = \nabla^2\Phi_N = - \frac{3}{2}\Omega_{m0}\mathcal{H}(\tau) \delta_{m}(\tau)= -\kappa \delta_{m},
\label{eq:poisson}
\end{equation}
where $\kappa = \frac{3}{2}\Omega_{m0}\mathcal{H}(\tau)$ and $\mathcal{H}$ beings the conformal Hubble expansion rate. 
In LPT, the above equation is generally solved by expanding the displacement vector perturbatively as $\bf{\Psi} = \bf{\Psi}^{(1)} +\bf{\Psi}^{(2)}+ ...$ where $\bf{\Psi}$ represents a curl free, gradient of a scalar field, $\phi^{(i)}$, $\bf{\Psi}^{(i)} = \bf{\nabla}_{q}\phi^{(i)}$. Similarly, expanding the density contrast into a perturbative series: $\delta_m(x) = \delta_m^{(1)}+ \delta_m^{(2)}+ ... = J^{-1}-1$ with the Jacobian of transformation, $J = \rm{Det}(\delta_{ij} +{\bf{\Psi}}_{i,j})$, we can subsequently equate each order of it to the displacement-field vector order as follow:
\begin{equation}
\delta_{m}^{(1)} = - \bf{\Psi}^{(1)}_{i,i},
\end{equation}
\begin{equation}
\delta_{m}^{(2)} = - \bf{\Psi}^{(2)}_{i,i} +\frac{1}{2}\left((\bf{\Psi}^{(1)}_{i,i})^2 +(\bf{\Psi}^{(1)}_{i,j})^2\right),
\label{eq:delta2}
\end{equation}
where $\bf{\Psi}_{i,j}=\partial \bf{\Psi}_{i}/\partial q_{j}$ and the divergence w.r.t $x$ is changed into $q$ via the Jacobian transformation,$\nabla_{x_i} =(\delta_{ij}+{\bf{\Psi}}_{i,j})^{-1}\nabla_{\bf q_j}$. 
At the first order we have from equation(\ref{eq:poisson}):
\begin{equation}
\frac{d^2 {\bf\Psi}^{(1)}_{i,i}({\bf q},\tau)}{d\tau^2} = -\kappa \delta_{m}^{(1)}({\bf q},\tau),    
\end{equation}
which is equivalent to 
\begin{equation}
\left(\frac{d^2}{d\tau^2}-\kappa\right) \nabla^2 \phi^{(1)}({\bf q},\tau) =0,
\label{eq:1st_LPT}
\end{equation}
where $\phi^{(1)}({\bf q},\tau)$ is factorized into time dependent normalized growth factor,$D^{(1)}(\tau)$ multiply by $\phi^{(1)}({\bf q},\tau_{in})$ and $\phi^{(1)}({\bf q},\tau_{in})$ is the initial condition field come from $\delta^{(1)}({\bf q},\tau_{in})$. Thus, the growth factor $D^{(1)}(\tau)$ follows the equation:
\begin{equation}
\left(\frac{d^2}{d\tau^2}-\kappa\right) D^{(1)} =0,
\end{equation}
and it can  be solved easily by assuming the initial conditions according to the growing mode of the matter dominated ``Einstien de - Sitter'' Universe, $D_{in}^{(1)} =1$ and $\frac{dD^{(1)}_{in}}{d\tau}=\left(\frac{1}{a}\frac{da}{d\tau}\right)_{\tau =\tau_{in}}$. Note that it is the same solution we obtained from eq(\ref{eq:2lptDE1}).  So, the first order displacement field at any time is simply given by ${\bf \Psi}^{(1)}(q,\tau)= D^{(1)}(\tau){\bf \Psi}^{(1)}(q,\tau_{i})$, i.e once the displacement field at the initial time is computed,  $\bf{\Psi}^{(1)}$  in simulation, we can track for it anytime by multiplying it by the growth factor.
Similarly, using the equations (\ref{eq:delta2}) and (\ref{eq:poisson}), the second order Lagrangian perturbation takes the form of 
\begin{equation}
\left(\frac{d^2}{d\tau^2} -  \kappa \right)\nabla^2\phi^{(2)} = -\frac{\kappa}{2}\left[(\nabla^2\phi^{(1)})^2 - (\nabla_{i}\nabla_{j}\phi^{(1)})^2\right],
\label{eq:2nd_LPT}
\end{equation}
with $\phi^{(2)}(q,\tau) = D^{(2)}(\tau)\phi^{(2)}(q,\tau_{in})$ and the second order growth factor $D^{(2)}$ satisfies 
\begin{equation}
\frac{d^2 D^{(2)}}{d\tau^2}-\kappa D^{(2)} = -\kappa D^{(1)2}.
\label{eq:D2}
\end{equation}
 We again provide the initial conditions according the Einstein-de Sitter Universe; $D^{(2)}_{in} = -\frac{3}{7}$ and $\frac{dD_{in}^{(2)}}{d\tau} = - \frac{6}{7}\left(\frac{1}{a}\frac{da}{d\tau}\right)_{in}$ and the solution is nothing but the solution of (\ref{eq:2lptDE2}). The second order initial field $\phi^{(2)}(q,\tau_{in})$ is given by
 \begin{equation}
\nabla^{2}\phi^{(2)} = \frac{1}{2}\left[(\nabla^2\phi^{(1)})^2 - (\nabla_{i}\nabla_{j}\phi^{(1)})^2\right]. 
 \end{equation}
Hence, in LPT, the equation (\ref{eq:LPT_position}) reduces to 
\begin{equation}
{\bf x} = {\bf q}- D^{(1)}{\bf\nabla_{q}}\phi^{(1)}+D^{(2)}{\bf \nabla_q}\phi^{(2)}, 
\label{eq:LPT_position_pot}
\end{equation}
and the velocity field, $v$ is governed by
\begin{equation}
{\bf v} =\frac{d{\bf x}}{d\tau} = -D^{(1)}f^{(1)}\mathcal{H} {\bf \nabla_{q}}\phi^{(1)}+D^{(2)}f^{(2)}\mathcal{H}{\bf \nabla_{q}}\phi^{(2)},
\end{equation}
where $f^{(1,2)}$ represent the growth rate $f^{(1,2)} = \frac{d\ln D^{(1,2)}}{d\ln a}$.

For our dynamical DE models, $D^{(1,2)}$ and $f^{(1,2)}$ are provided by solving the equations (\ref{eq:2lptDE1}) and (\ref{eq:2lptDE2}) along with their Hubble expansions. The main differences observed among each others and from $\Lambda$CDM have been discussed in above subsection (\ref{sec:linear}).


\subsubsection{COLA Approach}
In the COLA method, the equation of motion is solved in the frame of reference comoving with the particles in Lagrangian space. So, the residual displacement, $\Psi_{res}$ after the COLA approach applied is basically the one where the linear (Zel'dovich) and the quasi-linear 2LPT displacement fields,  ($\Psi^{(1)}$,$\Psi^{(2)}$) are being subtracted from the full non-linear displacement that each particle would experience: 
\begin{equation}
\Psi_{res} = \Psi - D^{(1)}\Psi^{(1)} - D^{(2)}\Psi^{(2)}.    
\end{equation}
Hence, the equation of motion(\ref{eq:geodesic2}) takes the form:
\begin{equation}
T^2[\Psi_{res}] = -\nabla\Phi_{N}- T^2[D^{(1)}]\Psi^{(1)} - T^2[D^{(2)}]\Psi^{(2)} 
\end{equation}
where $T^2[X] = \frac{d^2X}{d\tau^2} +\mathcal{H}\frac{dX}{d\tau}$. The residual displacement, $\Psi_{res}$ is allowed to compute by the usual Particle-Mesh scheme.

For more details about the implementation of the method and the way the time integration has been discretized on the operator, $T$, we refer the reader to \cite{Tassev:2015mia, Howlett:2015hfa}. 
Note that in the limit of large number of time-steps, it recovers the same as the standard approach on both the large and small scales.
Thus, in the COLA approach the run time is directly proportional to the number of time steps we set to run the simulation. The previous studies of the COLA approach have already shown that just by 10 steps, it could converge to full-N body simulation up to $2\%$ for $k = 1h/Mpc$. So, thus, the convergent rate increases as we increase the number of time steps. Pointed out that our motive of the work is not test the accuracy of the code, rather to study the effect of dynamical DE models.


\subsection{Simulation set up}

For this work, we use the Planck's \cite{Ade2015} cosmology: $\Omega_{m0}=0. 3089, h =0.6774, n_s=0.9667, \sigma_8=0.8159$ and the model parameters are set according to the best fit values from \cite{Jaber:2017bpx}. Specifically, the CPL-limit is defined with $w_{0} = -0.92, w_{i} =-0.99, z_{T} =1 =q$ and SEoS models by $w_{0} =-0.92, w_{i} =-0.99, z_{T}=0.28, q=9.98$. Recall that SEoS-I and SEoS-II models are differentiated by the fact that SEoS-II is run with $\Omega_{m0}=0.334$ and $ h=0.7322$. The N-body simulations are run on the box of side length: $L_{\rm box}=1024 h^{-1}{\rm Mpc}$ (L1024) with the number of dark matter particles, $N_{p} = 1024^3$ on mesh, $N_{\rm mesh}= 1024$ for FFT. These lead to the mass of the particles: $8.57198 \times 10^{10}{\rm M_{\odot}}h^{-1}$ for $\Omega_{m0}=0.3089$ and $9.2685 \times 10^{10} {\rm M_{\odot}}h^{-1}$ for $\Omega_{m0} = 0.334$ respectively. 


The theoretical primordial matter power spectra input to the simulations are taken from \texttt{CAMB} \footnote{https://camb.info/}. The initial particles distributions are generated by using the 2LPTic initial condition code \cite{Crocce:2006ve}\footnote{http://cosmo.nyu.edu/roman/2LPT/} at the initial redshift of $z_{in}=49$. Upon the modification of the COLA-code after cooperating the behavior of dynamical DE models, we run the simulation with 200 time steps between redshift $z_{in}=49$ till $z=0$ and take several snapshots at different redshift for each model. To suppress sampling variance on our results, we estimate the results by averaging 5 independent realizations, run with different random seeds. Our results will be focused at redshifts, $z=0, 0.28 $ and $1$. Keeping in mind, $z=0.28$ and $z = 1$ being the transition points ,$z_{T}$ for SEoS models and CPL-limit respectively. 
A brief summary of cosmological and the simulations parameters are listed out in tables (\ref{table:COLA_models}) and (\ref{table:simulation_COLA}). 

\section{Results}
\label{sec:results}
The effect of dynamical dark energy are often studied through both the dark matter particles and halo properties. Particularly, the dark matter power spectrum, velocity power spectrum, halo abundance and halos-clustering at different redshifts are widely analyzed. For our case, we study the matter power spectra calculated using the \texttt{SimplePofk} code \footnote{http://ascl.net/1110.017} \cite{Colombi2011} where the density field is estimated using a Cloud-in-Cell mass assignment method on a cubic grid with the same resolution as the Particle-Mesh grid used for the integration of the N-body system. The halo catalogs are generated using a publicly available phase-space temporal halo finder called \texttt{rockstar} \footnote{https://bitbucket.org/gfcstanford/rockstar} \cite{Behroozi2013}. The \texttt{rockstar} halo-finder is set to find  the halos with the minimum of at least $20$ particles to consist as a halo which leads to the minimum mass of the halo of $\sim 2 \times 10^{12}{M_{\odot}}h^{-1}$. Later we study the effect on the dark matter halos clustering by calculating the two point correlation function at real space. 

\subsection{Non-linear matter power-spectrum} 

\begin{figure*}
     \centering
     \begin{tabular}{cc}
        \includegraphics[width=1\textwidth,angle=0]{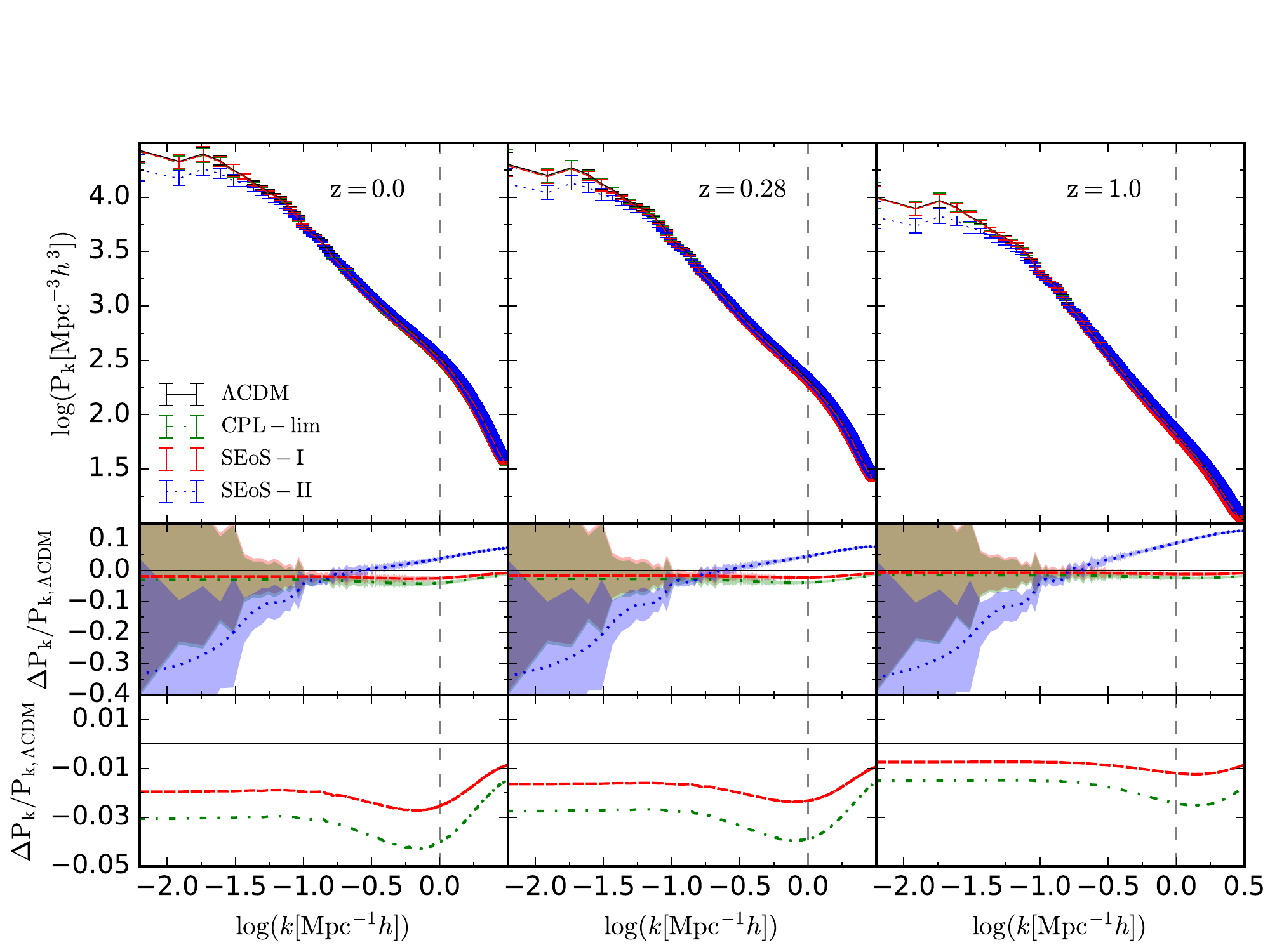}   

 \end{tabular}

\caption{{\it Upper Panels:} The dark matter power spectra for all the models at redshifts: $z = 0, 0.28, \& 1$. We follow the same color coding to represent the corresponding models as before. {\it Middle panels}: The relative difference of all models w.r.t $\Lambda$CDM. Shaded portions represent one sigma standard deviation propagated from 5 realizations run with the different random seeds. {\it Lower panels}: Focusing on the relative differences of SEoS-I and CPL-lim w.r.t $\Lambda$CDM. The vertical line provides the limit beyond which $P_{k}$ can not be trusted. 
}
\label{fig:pk_nonlinear}
\end{figure*}


\label{sec:pk} 
 In figure (\ref{fig:pk_nonlinear}) the non-linear matter power spectra, $P_{k}(k,z)$ of all the models at $ z=0, 0.28 $ and $1.0$ are plotted based on the simulation of ${\rm L1024}$. In each panel, solid black, green dotted-dash, red dashed and blue dotted lines correspond to $\Lambda$CDM, SEoS-I, CPL-lim and SEoS-II models respectively. The relative differences with respect to $\Lambda$CDM for all the models are shown in the middle panels. The shaded portions represent the error propagated from the $1\sigma$ standard deviations of 5 realizations, run with the different initial random seeds. The bottom panels are to highlight the difference of SEoS-I and CPL-lim from $\Lambda$CDM.  

The $P_{k}$ of SEoS-I(red dashed) and CPL-lim(blue dotted) are deficits in power of $\sim 2\%$ and $\sim 3\%$ respectively throughout all k-scales w.r.t $\Lambda$CDM, with a slight increase at the non-linear regimes (see bottom panels of figure (\ref{fig:pk_nonlinear})) that reaches to $\sim 3-4\%$. Note that the maximum deviation observed in the linear growth $\delta^{(1,2)}$ for these models are within $\sim 2-3\%$ at $z = 0, 0.28, 1$. This points out that the impact of dynamical DE is relatively small both in linear and non-linear regimes.
However, the overall deficits in $P_{k}$ values for SEoS-I and CPL-lim are presenting the impact of different DE in the DM growth. Explicitly, the effect of having less negative values of $w(a)$ for both SEoS-I and CPL-lim compare to $\Lambda$CDM 
leads the models to evolve with slightly larger expansion, $H(a)$ and ends up to enter the DE dominated epoch relatively earlier then $\Lambda$CDM. This results in slowing down the growth of structures even more then in the $\Lambda$CDM universe. 
The trends of reduced in deviation of $P_{k}$ with the increase of redshifts is again showing the effect of DE behaviors, as a consequent of their expansion rates. In particular, $2 \%$ and $3 \%$ differences at $z = 0$ reduce to $1 \%$ and $2 \%$ at $z = 1.0$ for SEoS-I and CPL-lim respectively.

On other hand, the $P(k)$ of SEoS-II can be understood by recalling the behavior of linear power spectrum, $P_{k,lin}$ of $\Lambda$CDM  on varying $\Omega_{m0}$ and $H_{0}$. For instance, looking at $P_{k,lin}$ of $\Lambda$CDM at $\Omega_{m0} =0.3089$ and $\Omega_{m0}=0.334$ for the same $H_{0}$, we found that $P_{k,lin}$ of $\Omega_{m0}=0.334$ have deficits in power at low $k$ modes (linear regime), then trends to increase and ends up into the larger in power at high $k$ modes (non-linear regime). Larger the value of $\Omega_{m0}$, the more deviation we observed in $P_{k,lin}$, depicting the direct proportionality of $P(k)$ with $\Omega_{m0}$ both in linear and non-linear regimes (see for the similar effect in figure \ref{app:Pk_initial}).
Since $H_{0}$ doesn't change with time, the difference in $P_{k}$ at different redshifts are not effected by $H_{0}$.

Even though, the effect of DE in general is larger at the linear regime over the non-linear, but in case of SEoS-II where the matter contained, $\Omega_{m0} = 0.334$ is much higher then other models including $\Lambda$CDM, i.e. $\Omega_{m0} = 0.3089$, the structures formed in such environments are more dense (see $\delta^{(1,2)}$ in figure (\ref{fig:deltas})), thus, subsequently leads to low values in $P_{k}$ at low $k$ modes. The impact of $\Omega_{m0} = 0.334$ and $H_{0} = 0.7332$ results in more clustering at the non-linear regimes. More specifically, our result in figure (\ref{fig:pk_nonlinear}) shows a difference at-most of $\sim 30\%$ at $k= 0.001$ between the SEoS-II model and $\Lambda$CDM while the difference reduces up to $\sim 10\%$ at the non-linear regime (say $k=10$). The slight reduced in power of $P_{k}$ of SEoS-II on going from $z = 1$ to $z = 0$ in high k modes, is due to DE effect that we observed in $H(a)$ figure(\ref{figure:hubble}). To this end we found that the effect of having a steep transition in the DE models is difficult to detect in $P_{k}$ infront of overall impact of DE.
As expected from the background studies, SEOS-I turns out to be the closest one to $\Lambda$CDM, followed by CPL-lim and SEoS-II


\begin{figure*}
     \centering
      \includegraphics[width=1\textwidth,angle=0]{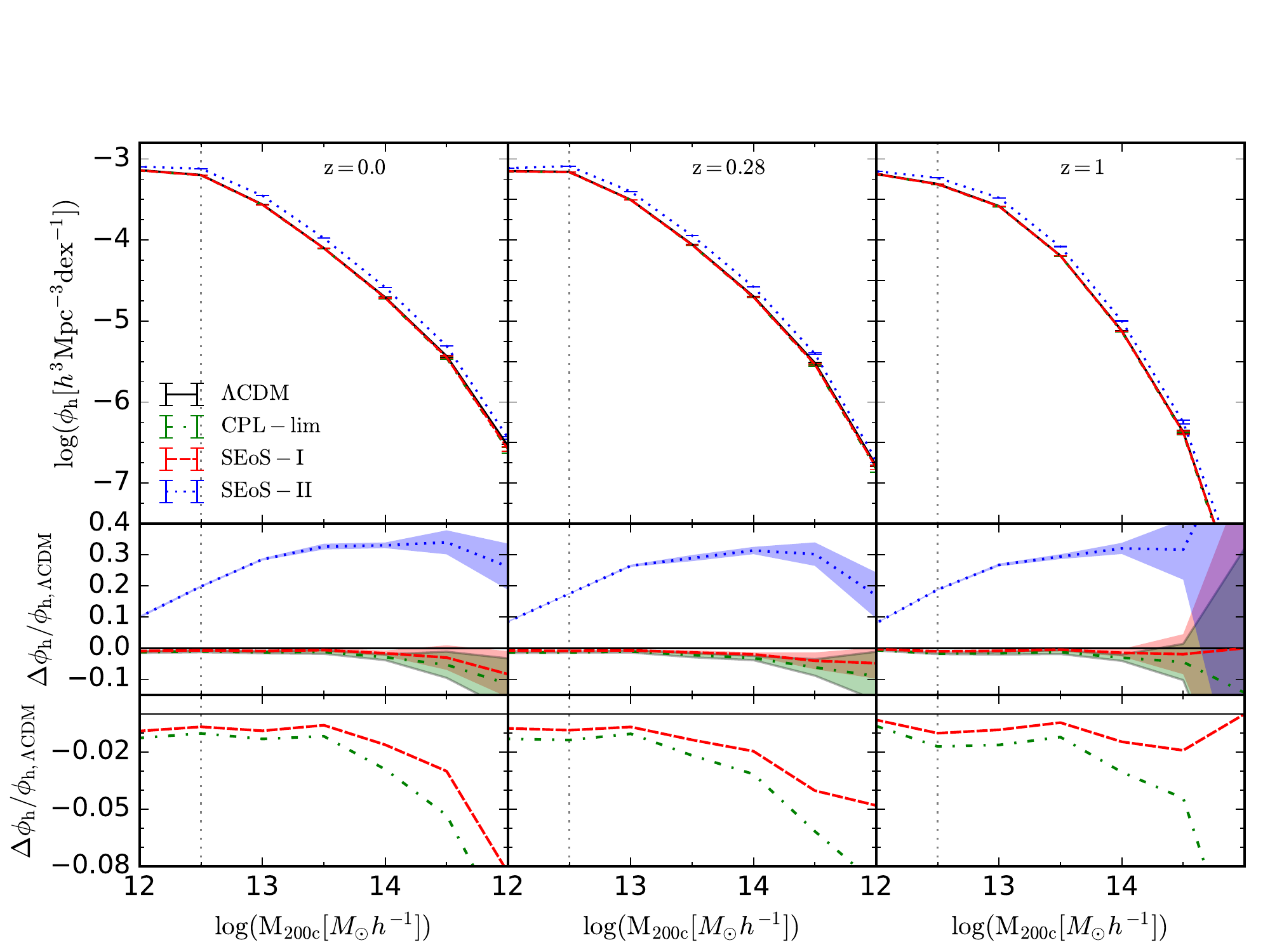}

\caption{{\it Upper Panels:} Differential halo mass functions for all the models at redshifts, $z = 0, 0.28~\& 1.0$. {\it Second panels}: The difference with respect to $\Lambda$CDM along with one sigma standard deviation calculated from 5 realizations run under different random seeds in shaded regions.{\it Third panels}: Zoom-in plot of the second panel to higlight the differences between the CPL-lim and SEoS-I.
}
\label{fig:HFM}
\end{figure*}

\subsection{Halo mass function}
\label{sec:massfunction}



Depending upon the behaviors of DE, the structure we observed can be more or less clustered, subsequently, will take lesser or longer time to form a gravitational bound structure. Such effect are studied via the Halo mass function (HMF). HFM measures the number of halos per unit volume under certain range of mass bin, $M$ to $M+dM$ at some particular redshift.  We use the halo mass definition of $M_{200} \equiv \frac{4\pi}{3}200\rho_{c}R^3_{200}$ which corresponds to halos enclosing $200$ times the critical density of our Universe and their corresponding radius $R_{200}$. 

Figure (\ref{fig:HFM}) shows the mean differential HMF of 5 realisations for each model at redshifts, $z = 0, 0.28$ and $ 1$. As expected SEoS-I has the closest halo abundance to $\Lambda$CDM, followed by CPL-lim and SEoS-II. SEoS-II beings a universe with $\Omega_{m0}=0.334$, the number density of DM halos is expected to be larger then others. That is what we found in HMF(see {\it Middle panels} of figure (\ref{fig:HFM})) throughout whole mass ranges and at all redshift considered. The difference of $\sim 30\%$ in HMF is observed between SEoS-II and $\Lambda$CDM in the mass range of $10^{13} M_{\odot}h^{-1} < M_{200}< 10^{15} M_{\odot}h^{-1}$. The difference is slightly reduced at both low and high mass ends. This reduced in HMF at the high mass end is probably because of the difference in DE behaviors. However, the most prominent effect is coming from the difference in the global cosmological parameters.

The {\it bottom panels} of figure (\ref{fig:HFM}) highlights the main different in HMF of SEoS-I and CPL-lim from $\Lambda$CDM. Noting that they share the same cosmological parameters: $\Omega_{m0}=0.3089$ and $H_{0} = 67.74$, the main different is basically driven by the different in dynamic of DE. We observe that SEoS-I and CPL-lim have lower number of halos compare to $\Lambda$CDM at all mass scales and redshift.  
The shower growth rate of matter densities in SEoS-I and CPL-lim w.r.t $\Lambda$CDM results into the lower number of halos. One can certainly confirm from this figure that the chance of distinguishing between these DE models lie in the high mass halos. Particularly, looking at mass scale of $M_{200}=10^{14.5} M_{\odot}h^{-1}$, a  significant difference of $\sim 3\%$ to $\sim 5\%$ are observed at $z = 0$ in SEoS-I and CPL-lim from $\Lambda$CDM respectively. However, going toward the high mass limits comes with the cost of uncertainty.
The difference in HMF reduces as we go from low to high redshift for both SEoS-I and CPL-lim. We argue that by increasing the number of particles in the simulation setup might help in reducing the uncertain level and increase the possibility to distinguish between the models. 


\begin{figure*}
     \centering
     \includegraphics[width=1\textwidth,angle=0]{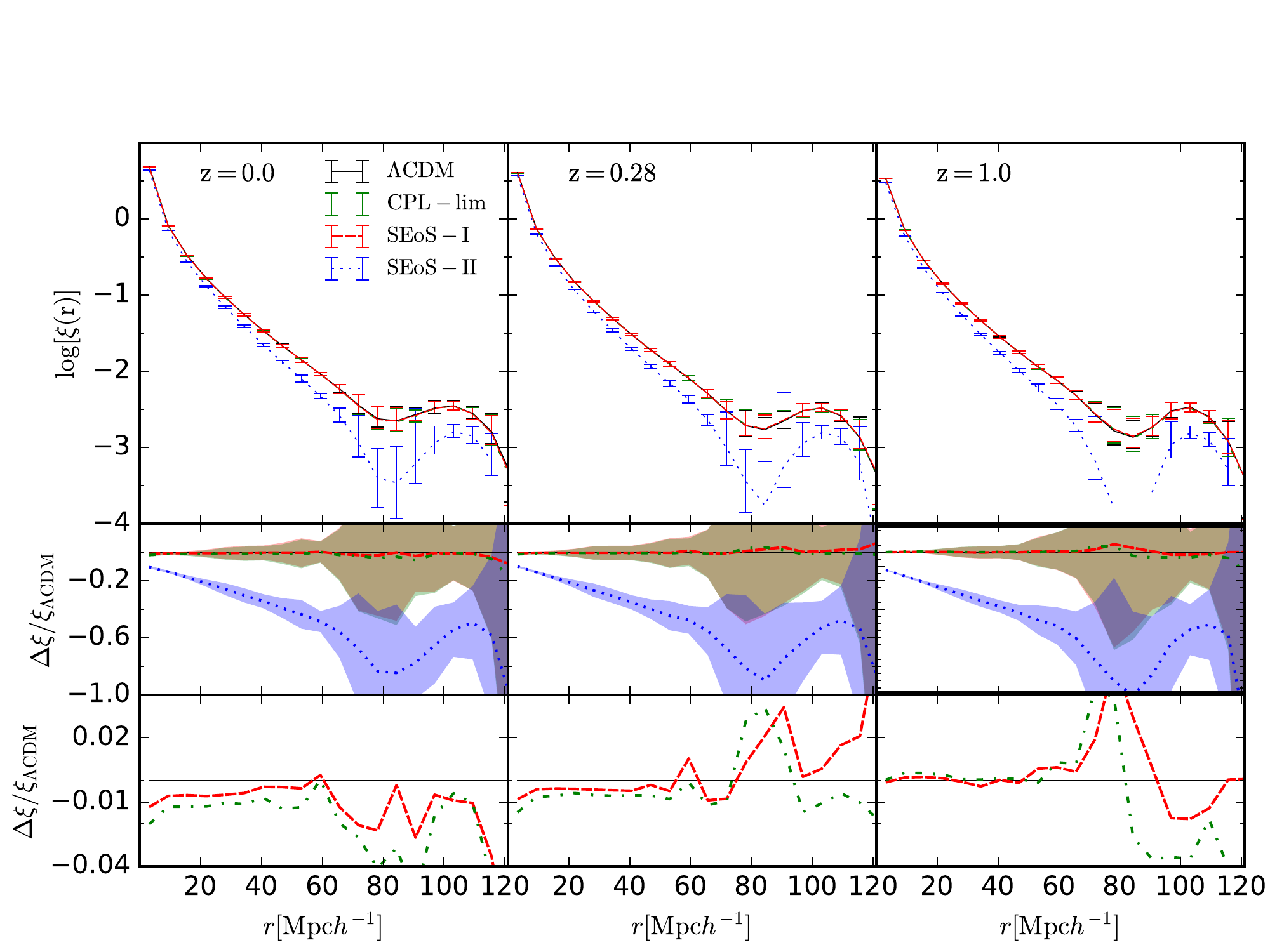}


\caption{2PCF for all the considered models at redshifts, $z = 0, 0.28~\& 1.0$ after selecting Halos with more than 50 particles. The respective difference w.r.t $\Lambda$CDM are shown in {\it middle} and {\it lower} panels. Shaded portions represent one $\sigma$ standard deviation calculated from 5 realizations run under different random seeds. Note: A significant deviation is observed for SEoS-II model from the $\Lambda$CDM at all redshifts.}
\label{fig:2PCF}
\end{figure*}

\subsection{Two point correlation function}

The two point correlations function remains one of the strongest tools to study the clustering of large scale structure either through galaxies or halos. For this work, we analyse DM halos clustering in real space for all the models. Interestingly, our simulation setting of $L_{\rm box}$= $1024 h^{-1}{\rm Mpc}$ and $Np=1024^3$, might allow us to trace up to the Baryon Acoustic Oscillation(BAO) scales.

The measurement of the two point correlation function(2PCF) between the halos are carried out by using the  code called Correlation Utilities and Two-point Estimates (CUTE) \cite{cutecode}\footnote{https://github.com/damonge/CUTE}. The correlation function we measured is estimated by the function:
\begin{equation}
\xi\left(r\right)=\frac{V_{\rm box}\times D_{h}D_{h}\left(r\right)}{V_{\rm bin}\left(r\right)\times N_{h}^{2}}-1
\end{equation}
where $D_{h}D_{h}$ is the number of halo-halo pairs within a given separation bin of volume $V_{\rm bin}$, $N_{h}$ is the total number of halos in the box and $V_{\rm box}$ is the volume of the simulation box, i.e $V_{\rm box}=L_{\rm box}^3$. 20 linearly spaced bins are set up in the range of $0-125~h^{-1} {\rm Mpc}$. For a statistically homogeneous and isotropic field, the halos are only correlated according to their relative distance $r$, that is we assume for our case. Thus, we present the result of 2PCF as a function of $r$ in figure (\ref{fig:2PCF}) for $z=0,\,0.28,\,1.0$.
Note that our results are based on the average over 5 realizations of simulations we run for each models.
Similar to figures (\ref{fig:pk_nonlinear}) and (\ref{fig:HFM}), the results are shown in three panels (\ref{fig:2PCF}). The upper, middle and lower panels are focused on the full 2PCF, relative difference of models w.r.t $\Lambda$CDM along with the error estimated from 5 realisations for each models and zoom in of middle plots to highlight the difference of SEoS-I and CPL-lim from $\Lambda$CDM respectively. 

From the upper panels of (\ref{fig:2PCF}), we observe the BAO feature at the scale around $\sim 100 h^{-1} {\rm Mpc}$ in all models. The standard behaviors of the BAO feature are recovered including the BAO bumb is becoming wider when it goes from high to low redshift, even with the limitation in resolution at that scales.

A significant different in the 2PCF is observed (see in the middle panels of (\ref{fig:2PCF})) between SEoS-II and $\Lambda$CDM. As discussed before the difference between SEoS-II and $\Lambda$CDM is mainly governed by the difference in global cosmological parameters i.e. $\Omega_{m0}=0.334$ and $H_{0} = 73.32$ in contrast to $\Lambda$CDM: $\Omega_{m0}=0.3089$ and $H_{0} = 67.74$. Thus, we draw the following conclusion, the low clustering revealed in the 2PCF of SEoS-II is because the model with the higher DM content leads the structures to grow faster (see figures \ref{fig:deltas} and \ref{fig:f_growth}) and ends up into more dense halos but widely spread out. In addition, DE dominated epoch comes later in this model compare to $\Lambda$CDM, hence the structures tend to virialise earlier. This result also shows us the sensibility of halo clustering on the cosmological parameters. The difference in 2PCF of SEoS-II over $\Lambda$CDM reduces from $\sim 70\%$ at $r = 70 h^{-1} {\rm Mpc}$ to $\sim 20\%$ at $r = 20 h^{-1} {\rm Mpc}$. Note that the errors become relatively high after the scale, $r = 60 h^{-1} {\rm Mpc}$.


 The bottom panels of (\ref{fig:2PCF}) show that CPL-lim and SEoS-I models remain in around $\sim 1-2\%$ difference in halo clustering up to the scales where the scatter is low, say $r = 60 h^{-1} {\rm Mpc}$. Hence, we conclude that it is challenging to trace the effect of dynamical DE using the halo clustering when we have a $\sim 2\%$ difference in the background expansion rate and the linear growth. We argue that including a larger number of simulations and increasing their accuracy with a larger number of particles and spatial resolution will improve further on this test. 

\section{Summary and Conclusions}
\label{sec:conclusions}



 In this paper, we presented a set of N-body simulations designed in the framework of dynamical DE models and analyzed their impacts in the structure formation. Particularly, we studied the effect that a rapid dilution in the energy density of DE has in the growth of structure at cosmic scales. To this end we assumed DE models with the equation of state, $w(z) = w_0 + w_i\frac{(z/z_T)^q}{1+(z/z_T)^q}$, where a steep transition from a high negative EoS phase to a low negative phase is controlled by the large values of $q$, at a given epoch, set by $z_T$.
 
 Specifically, we compared a DE model with the steep transition at late time, $z_T$=0.28 with $q=9.98$ versus the same model with a smoother transition at a particular redshift, $z_T=1$ with $q=1$. The later choice of parameters ended into a Chevallier-Polarski-Linder parameterization (CPL-lim) like model \cite{Chevallier2001,Linder2003}. The former refered to the SEoS model, that we further divided into two sets: SEoS-I and SEoS-II depending on the cosmological parameters $\Omega_{m0}$ and $H_{0}$ used. The SEoS-I is based on the Planck`s results (P15)\cite{Ade2016}: $\Omega_{m0}=0.3089$ and $ H_{0}=67.74$ as in $\Lambda$CDM and CPL-lim. The SEoS-II model considered $\Omega_{m0}= 0.334$, $H_{0} = 73.22$, the best fit values obtained in \cite{delaMacorra:2015aqf} along with the SEoS parameters. This SEoS-II is claimed to be the best fitted
model according to \cite{delaMacorra:2015aqf}.


The structure formation under such DE scenarios were studied upon a modification of the approximated N-body simulation code called L-PICOLA \cite{Howlett:2015hfa}.   
The DE is implemented via the expansion rates and DM perturbation solutions into the Lagrangian perturbations method employed in the code.   
The simulations assume a Box length: $L_{\rm box}=1024 h^{-1}{\rm Mpc}$ and a number of dark matter particles, $N_{p} = 1024^3$ on a mesh, $N_{\rm mesh}= 1024$. In total, we ran 20 simulations, 5 realisations for each model to estimate the cosmic variance errors. Using the output of simulations, we analysised the non-linear DM power spectrum, $P_{k}$, the number density of the gravitational bound halos and halo clustering using the two point correlation function. We discussed our results at 3 redshifts, $z = 0, 0.28, 1$, considering $z = 0.28$ and $z = 1$ being the transition redshifts for SEoS models and CPL-lim, respectively. 

The main results and conclusions are as follows:
\begin{itemize}
\item {
The calculated non-linear power spectra, $P_{k}$ of SEoS-I and CPL-lim models found deficits in power w.r.t $\Lambda$CDM throughout all $k$-modes. Such behaviour is expected due to their dynamics of DE i.e overall less negative values of EoS and relatively slightly larger hubble expansion than $\Lambda$CDM trigger a slowdown in the growth of structure. This result is consistent with the linear matter perturbations studies. The differences remain within $\sim 2\%$ and $\sim 3\%$ throughout all $k$ scales at $z = 0$ for SEoS-I and CPL-lim, with a slight increase at non-linear regimes. 
The reduced in deviations from $\Lambda$CDM with the increase of redshift for both cases is also showing the effect of DE. As such no significant imprint of the steep transition in DE field is observed in the structure formation.

The $P_{k}$ of SEoS-II showed up relatively different behaviors. Specifically, a deficit in power at the linear scales and the difference reduces as it goes from linear to non-linear regimes. Such behavior in SEoS-II is expected for the universe governed by the high DM contained, $\Omega_{m0} = 0.334$ and $H_{0} = 72.23$ over other models with $\Omega_{m0} = 0.3089$ and $H_{0} = 67.74$. Thus, we conclude that the behavior of DE in SEoS-II model is subdominant upon the variation of $\Omega_{m0}$ and $H_{0}$.}

\item We discussed the differential halo mass functions(HMF) being effected by the behavior of DE models. The results of low HMF observed in SEoS-I and CPL-lim over $\Lambda$CDM at all mass scales is according to our prediction. The lower growth rate of matter densities in SEoS-I and CPL-lim w.r.t $\Lambda$CDM results into less number of bound halos. The deviation of SEoS-I and CPL-lim reduces with the increase of redshift. The chances of distinguish between these models from $\Lambda$CDM increases at the high mass ends. 
The SEoS-II model being governed by $\Omega_{m0} = 0.334$, have higher HMF as predicted and reached upto $\sim 30\%$ deviation from $\Lambda$CDM at $ 10^{14.5}M_{\odot}h^{-1}$ at the considered redshifts.

\item {We quantified the 2PCFs of SEoS-I and CPL-lim models and both remain within $\sim 1-2\%$ difference w.r.t $\Lambda$CDM upto $r = 60 {\rm Mpc} h^{-1}$, at larger scales the accuracy of numerical experiments make challenging to state robust conclusions, but they behave in a consistent way. On other hand, SEoS-II formerly considered as the best fit model \cite{Jaber:2017bpx}, considering background and linear growth, shows a significant difference throughout all $r$ scales. Such difference reduces going from large to small $r$ scales. The deviation reached upto $\sim 60\%$ at $r = 60 h^{-1} {\rm Mpc}$. We conclude that including the non-linear growth is critical in order to asses the viability of DE models.}

\item {Although not surprisingly for approximated simulations \cite{2016JCAP05051S}, results of halo 2PCF and HMF might be biased by the ability of experiments to resolve halo mass and size, in opposition to $P_{k}$ that uses the particle distribution. However the general trend of results is consistent across all tests.}

\item {We pointed that having a tool to study LSS based on the N-body simulation that run at low computational costs and simultaneously recovers the LSS accurately under the dynamical DE scenarios based on L-PICOLA\cite{Howlett:2015hfa} would be widely useful to understand the nature of DE. Particularly, to provide the constrain on DE model parameters after running multiple sets of simulations on the wide ranges of EoS and the cosmological parameters using the clustering measurements.}

\end{itemize}

On conclusion, we state that our results from the non-linear structure formation are consistent with the behaviour of DE, their background and linear perturbation theories. Differences found in SEoS-I and CPL-lim models are directly driven by the DE dynamics. The deviation observed in SEoS-II is entangled with the effect of cosmological parameters and DE behavior. Recalling that SEoS-II is the best fitted model of SEoS DE according to \cite{Jaber:2017bpx} where \cite{Jaber:2017bpx} used the background dependence observabls such as the latest local Hubble constant measurement (\cite{Riess:2016jrr}) and the compressed CMB likelihood from Planck collaboration (\cite{Ade2015, mukherjee, planck15DE}) to constraint these models. Our finding of large deviation in SEoS-II from $\Lambda$CDM in the non-linear structure formation studies is showing a hint to go beyond the linear theories to constraint DE models in general.
 
Note that our work was mainly focused on some particular DE parametrization, but the code is easily extended to other dynamical DE models as long as the DE field remains homogeneous. Even though, there is still room to further test SEoS models by increasing simulation number and accuracy. We are in progress of including other dynamical DE models such as the tachyon DE field and interacting DM-DE models in up-coming projects.  An strategy like the one presented in this paper is particularly appealing under the light of upcoming high accuracy galaxy surveys like DESI.

\appendix
\section{The linear power spectrum}
\label{app:Pk_initial}
\begin{figure}
\centering 
\includegraphics[width=0.6\textwidth,angle=0]{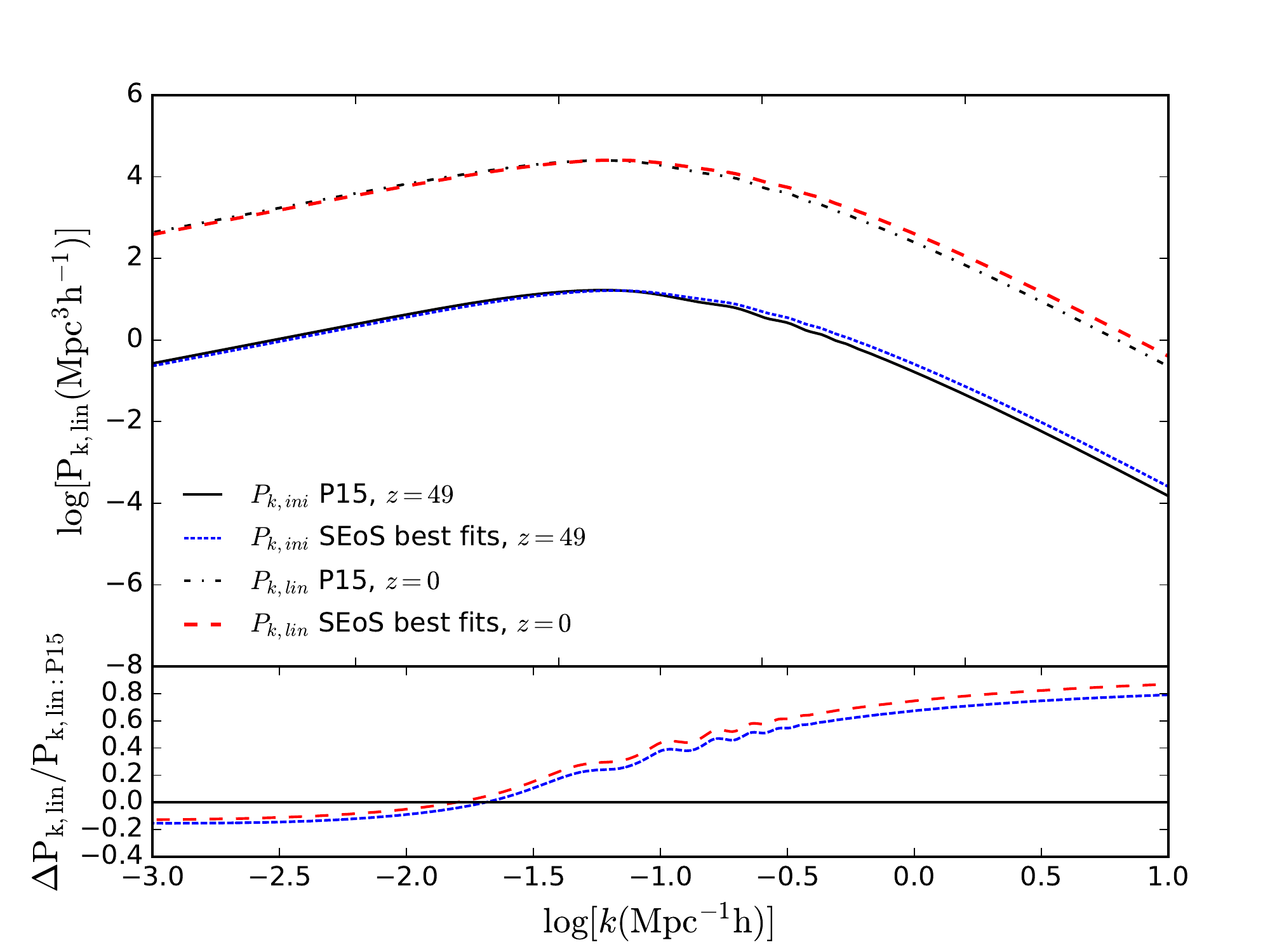}
\caption{The initial linear power spectrum from \texttt{CAMB} at redshift $z_{in}=49$, used for our simulations. The black solid and blue dotted lines correspond to the power spectra with P15 cosmological values $\Omega_{m0} = 0.308$ and SEoS-II best fit values, $\Omega_{m0} = 0.334$ respectively. Their corresponding CAMB output at $z=0$ are also included for comparison in red dashed and black dash-dotted lines respectively. 
\label{fig:pk_initial}}
\end{figure}

The initial linear power spectra, $P_{k,lin}$, at redshift $z_{in}=49$ from \texttt{CAMB}, used in the simulations are shown in figure (\ref{fig:pk_initial}). The black solid and blue dotted lines correspond to the power spectra obtained with P15 ($\Omega_{m0} = 0.308$ and $H_{0} =0.6774$) and SEoS-II case: ($\Omega_{m0} = 0.334$ and $H_{0} =73.22$) respectively, see for other parameters in the table 1. For comparison, we also cooperated $P_{k,lin}$ at $z = 0$ from the CAMB in black dash-dotted and red dashed lines for $\Lambda$CDM and SEoS-II cases respectively. We found that even at the initial power spectrum at $z_{in}=49$ there is an enhance in power at the high $k$ modes compare to inter-mediate scales, followed by the deficit in power in the low $k$ modes. The similar behavior is observed at $z = 0$ with slightly increase in different at high $k$ scales. These differences are expected and mainly due to the different in dark matter contained at the present time.

\section*{Acknowledgements}
NCDevi thanks Hans Winther and Marc Manera for the fruitful discussions and feedback. 
NCDevi acknowledges support from a DGAPA-UNAM post-doctoral fellowship and CONACyT Fronteras de la Ciencia grant 281. 
NCDevi acknowledges support from the European Commission's Framework Programme 7, through the Marie Curie International Research Staff Exchange Scheme LACEGAL (PIRSES-GA-2010-269264) when main research of the work are obtained.
M. Jaber acknowledges the support of CONACYT graduate fellowship and that of the Polish Ministry of Science and Higher Education MNiSW grant DIR/WK/2018/12.  
Part of this work was supported by the ``A next-generation worldwide quantum sensor network with optical atomic clocks'' project, which is carried out within the TEAM IV programme of the Foundation for
 Polish Science co-financed by the European Union under the European Regional Development Fund.
OV and G Aguilar acknowledge support from UNAM PAPIIT grant IN112518. G. Aguilar thanks support from a CONACyT graduate fellowship.
M. Jaber and Axel de la Macorra acknowledge support from Project IN103518 PAPIIT-UNAM and A. De la Macorra of PASPA-DGAPA, UNAM. HV acknowledges support from IN101918 PAPIIT-UNAM Grant. The authors acknowledge DGTIC-UNAM for facilities on the Supercomputer MIZTLI at DGTIC-UNAM and also the ATOCATL Cluster Supercomputer of IA-UNAM.
The authors also thank Julio Cesar Clemente for his help in setting up software in these supercomputer.  

\bibliographystyle{JHEP}
\bibliography{refs}
\end{document}